\documentstyle[12pt]{article}

\font\sqi=cmssq8
\def\DR{\rm I\kern-1.45pt\rm R}
\def\DC{\kern2pt {\hbox{\sqi I}}\kern-4.2pt\rm C}
\begin{document}
\thispagestyle{empty}
{\hfill  Preprint JINR P2-98-108}\\

{\hfill hep-th/9805009}\vspace{2.5cm} \\
\begin{center}
{\Large The Hamiltonian Formalism for the Generalized Rigid Particles.}
 \vspace{1.5cm}
 \\
{\large A. Nersessian}\footnote{e-mail:nerses@thsun1.jinr.ru}
\vspace{0.5cm}\\
{\it Joint Institute for Nuclear Research,}\\
{\it Bogolyubov Laboratory of Theoretical Physics,}\\
{\it  Dubna, Moscow Region, 141980, Russia}\footnote{Permanent address.}\\
\vspace{0.5cm}
and\\
\vspace{0.5cm}
{\it International Center for Theoretical Physics,  Trieste, Italy}
 \end{center}
\vspace{1cm}
\begin{abstract}
The Hamiltonian formulation for the mechanical systems
with reparametri\-za\-ti\-on-invariant Lagrangians, depending on
the worldline external curvatures  is given,
 which is based on the  use of moving frame.

A complete sets of constraints are found for the
Lagrangians with quadratic dependence on  curvatures,
for the Lagrangians, proportional to an arbitrary  curvature, 
and  for the Lagrangians,
 linear on the first and second    curvatures.
\end{abstract}
\newpage
 \setcounter{equation}0
\section{Introduction}

As it is known, the curve in a $D$-dimensional space possesses
 $D - 1$ reparametrization invariants (external curvatures)
${\tilde k}_1,....,{\tilde k}_{D-1}$, which are the
functions of a natural parameter  ${\tilde s}$
(see, e.g. \cite{postnikov}).
Therefore, the general reparametrization-invariant mechanical action
in $D$- dimensional space can be defined as
  \begin{equation}
{\cal S} =\int F({\tilde k}_1,....,{\tilde k}_N)d{\tilde s},
\quad 0\leq N\leq D-1 .
\label{action}\end{equation}
Such systems we will  call by the models of generalized rigid
particles.

The mechanical systems depending on the first and second curvatures
 became rather intensively studied in the late eighties as toy models of
rigid strings and $(2+1)$-dimensional
 field theories with the Chern-Simon term \cite{polakov}.
 Before long, it became clear, mainly due to the studies of
M.Plyushchay that those systems are of independent interest.

 For instance, at
$D=(2+1)$, $F=c_0+c_1{\tilde k}_1+c_2{\tilde k}_2$, $c_0\neq 0$
 they  describe a massive relativistic anyon  \cite{misha};
 at  $D=(3+1)$, $F=c_0+c_1{\tilde k}_1$, $c_0\neq 0$, a massive
 relativistic boson  \cite{misha1};
 at $D=(3+1)$, $F=c{\tilde k}_1$, a massless particle with an
arbitrary (both integer and half-integer) helicity \cite{misha2}.
The system with $F=c_0+c_1{\tilde k}_1^2$ 
corresponds  to the effective action of 
relativistic  kink in the field of soliton \cite{kp}.

Recently, E.Ramos and J.Roca have found that the model with $F=c{\tilde k}_1$
 possesses the  $W_3-$  gauge symmetry \cite{rr}.
They have also shown in an implicit way that
a system with Lagrangian $F=c{\tilde k}_N$
possesses $N +1$ gauge  degrees of freedom, perhaps,
forming $W_{N+2}$-algebra \cite{rr2}.\\

{\it Which (iso)spinning particles 
are described by the models of  generalized rigid particles?}  \\

{\it Which gauge $W$-symmetries can be inherent in these models?}\\
To answer on these questions,
 one should know the dimension and structure of
phase spaces of the models under consideration, 
the  generators  of their
gauge symmetries, and then quantized the models.

First of all, this needs the Hamiltonian formulation
of the models with the action (\ref{action}). 
However, 
the Lagrangians of that  models  depend on  $(N+1)-$ order derivatives,
since
the external curvatures  are determined by 
the expressions
 $${\tilde k}_I({\tilde s})
=\frac{\sqrt{\det{\hat g}_{I+1}\det{\hat g}_{I-1}}}{\det{\hat g}_I },
 \quad \quad {(g_I)}_{ij}
\equiv {\bf x}_{(i)}{\bf x}_{(j)},\quad 
i,j=1,\ldots,I, $$
where ${\bf x}_{(i)}\equiv d^i{\bf x}({\tilde s})/(d{\tilde s})^i$.
  Thus, one should first replace the initial Lagrangian
by an equivalent  second  order one and then pass
to the Hamiltonian formalism in $2D(N+1)$-dimensional
 phase space.

In the latter transition, most authors neglect invariant
properties of Lagrangians, which 
state in   their dependence on external curvatures.
As a result, even the construction of the complete set of
 constraints requires tiring structureless calculations.
 For example, in the refered paper \cite{rr2}
the complete set of constraints
was constructed  only for $F=c{\tilde k}_2$, the latter being
essentially nonlinear.

In this paper, we suggest  more geometrical approach for
constructing the Hamiltonian formalism for the  models of generalized
rigid particles, which is based on the use of moving frame.

The resulting system is formulated  in terms
of the coordinates of the initial space ${\bf x}$,
 the components of moving frame ${\bf e}_i$,
and their conjugated momenta ${\bf p}$ and ${\bf p}_i$,  $i=1,\ldots,N$.
The Lagrangian multipliers  in the total Hamiltonian of the system
represents the external curvatures of trajectories.

We demonstrate efficiency of the presented formulation, constructing the
complete sets of constraints and Hamiltonians for models with the
following Lagrangians:

i) $F=\frac{1}{2}\sum_{i}^N b_i{\tilde k}^2_i
+\sum_{i=1}c_i{\tilde k}_i +c_0,\quad b_1b_2\ldots b_N\neq0$;
 This system is characterized by the lowest degeneracy and by absence
of the secondary constraints.

ii)  $F=c{\tilde k}_N$,  $\forall D, N<D$;
 The system is specified by the maximal (for a given $N$) degeneracy
and by $N+1$ gauge degree of freedom.
All the constraints arising here are quadratic.
Surprisingly, this model coincides with the model 
$N+1$-pointing  discreet string.

We show that systems with the Lagrangians, linear on external
curvatures possess the maximally possible set of (quadratic)
primary constraints.
When the Lagrangian contains the curvatures $k_a$, $a<N$,
the number of secondary constraints  and the gauge symmetries of Lagrangian
 is decreased.
To illustrate this phenomena, we present the  complete sets of
constraints for the thoroughly studied models with Lagrangians linear on 
first  and second curvatures.\\

Throughout the paper, we assume the signature of the initial
space $\DR^D$ to be  Euclidean,
which should not cause misunderstanding when passing to
the pseudo-Euclidean signature.

We use the following groups of indices:
$$i,j,k=1,\dots N;\quad a,b,c,d= 1,\ldots (N-1);\quad
\alpha,\beta=1,\ldots, (N-2);$$
and the notation:
$$F_{,i}\equiv {\partial F}/{\partial {\tilde k}_i},\quad
F_{ij}\equiv {\partial^2 F}/{\partial {\tilde k}_i \partial {\tilde k}_j}
$$
\begin{equation}
{\widetilde\phi}_{0.i}={\bf p}{\bf e}_i,\quad
{\widetilde\phi}_{i.j}={\bf p}_i{\bf e}_j- {\bf p}_j{\bf e}_i,\quad
{\widetilde\Phi}_{0.0}={\bf p}{\hat L}{\bf p},\quad
 {\widetilde\Phi}_{0.i}={\bf p}{\hat L}{\bf p}_i,\quad
  {\widetilde\Phi}_{i.j}={\bf p}_i{\hat L}{\bf p}_j,
\label{sectilde}\end{equation}
 where
$${\hat L}={\hat I}-\sum_{i=1}^{N}{\bf e}_i\otimes{\bf e}_i\;\;;
\quad\forall {\bf a},\;\; {\bf b}: {\bf a}\equiv a^A,
{\bf b}\equiv a^A,\quad {\bf a}{\bf b}=\sum_{A=1}^D a^A b^A.$$
 \setcounter{equation}0
\section{Frenet Formulae and  Legendre Transformation}
Consider the Hamiltonian formulation of 
the models of generalized rigid particle.

Let us rewrite the action (\ref{action}) as
  \begin{equation}
{\cal S} =\int F({k}_1/s,....,{k}_N/s)sd\tau
;\quad{\rm  where}\quad s\equiv |\frac{d{\bf x}}{d\tau}|,
\quad k_i\equiv s{\tilde k}_i,
\label{action1}\end{equation}
and introduce the moving frame $\{{\bf e}_\mu\}$ for
the trajectory of that system
\begin{equation}
 {\bf e}_\mu{\bf e}_\nu=\delta_{\mu\nu}\quad
{\dot{\bf x}}= s{\bf e}_1, \quad \mu=1,...,D.
\label{mf}\end{equation}
In these terms the external curvatures are defined by the Frenet
equations
\begin{equation}
{\dot{\bf e}}_\mu={k}_\mu{\bf e}_{\mu+1}-{ k}_{\mu -1}{\bf e}_{\mu-1},
\quad {\bf e}_0={\bf e}_{D+1}=0,
\label{ff}\end{equation}
 so
\begin{equation}
k_{\mu-1}={\dot{\bf e}}_{\mu-1}{\bf e}_{\mu},\quad
k^2_{\mu}={\dot{\bf e}}^2_{\mu} -k^2_{\mu-1} ,\quad 
{\dot{\bf e}_\mu {\bf e}_\nu}=0, \quad {\rm if}\quad |\mu-\nu| >1 .
\label{cf}\end{equation}
Note that $k_\mu\geq 0$, for $\mu=1,\ldots,(D-2)$, whereas $k_{D-1}$
(``torsion") can assume both positive and negative values.
If some $k_I\neq 0$, then $k_\mu\neq 0$, at
$\mu=1,\ldots,I-1$ . 
 Vice versa, if $k_I= 0$, then $k_\mu= 0$, at $\mu=I+1,\ldots, D-1$
  (see, e.g. \cite{postnikov}).

With the expressions  (\ref{mf}), (\ref{ff}), (\ref{cf}) at hands,
 we can replace the initial Lagrangian by the following one
\begin{eqnarray}
{\cal  L}&= F(k_1/s,..., k_N/s)s+{\bf p}({\bf{\dot x}}-s{\bf e}_1)
+\sum_{a}{\bf p}_a({\dot{\bf e}}_a- k_a{\bf e}_{a+1}+ k_{a-1}{\bf e}_{a-1})
&\nonumber\\
&-\sum_{i,j} d^{ij}\left({\bf e}_i{\bf e}_j-\delta_{ij}\right)-
{F}_{,N}\left(k_{N}-({\dot{\bf e}}^2_{N} -k^2_{N-1})^{1/2}\right) &
\label{lfo}\end{eqnarray}
where  $s, k_i, d^{ij}, {\bf p}_a, {\bf e}_i$
are  independent variables.\\

 Now we can perform the Legendre transformation for 
the Lagrangian (\ref{lfo}).\\
The variables  ${\bf p}_a$ represent the momenta
conjugated to
 ${\bf e}_a$,
whereas momenta, conjugated to $(s, k_a, d_{ij})$,
lead to the trivial constraints
\begin{equation}
 p^s\approx 0,\quad p^a\approx 0,\quad p^{ij}\approx 0.
\label{erunda}\end{equation}
Setting  $k_N\neq 0, F_{,N}\neq 0$  we find, that the momentum
conjugated to ${\bf e}_N$, is of the form
\begin{equation}
{\bf p}_N=F_{,N} \left({\dot{\bf e}}^2_N-k^2_{N-1}\right)^{-1/2}
{\bf{\dot e}}_N.
\label{pN}\end{equation}
So, taking into account  (\ref{cf}), we get the  constraints
\begin{eqnarray}
&\chi_{N.N}\equiv{\bf p}_N{\bf e}_{N}\approx 0,\quad
\chi_{N.\alpha}={\bf p}_N{\bf e}_{\alpha}\approx 0,
&\label{chi}\\
&\Phi_{N.N}\equiv{\bf p}^2_N-({\bf p}_N{\bf e}_{N-1})^2-F^2_{,N}
\approx{\widetilde\Phi}_{N.N}-F^2_{,N}\approx 0.
\label{PhiNN0}&\end{eqnarray}
Thus, after   Legendre transformation we  obtain the following
{\it total} Hamiltonian
\begin{equation}
{\cal H}_{T}= {\cal H} +
\lambda^{(s)}p_s+ \sum_a\lambda^{(k)}_a p^a+\sum_{ij}\lambda_{(d)ij}p^{ij},
\end{equation}
where
\begin{equation}
{\cal H}=s{\phi}_{0.1}+\sum_{a}k_a{\phi}_{a.a+1}+
\lambda\Phi_{N.N}+ \sum_{i,j} d^{ij}u_{ij}
+ \sum_{\alpha}\lambda_\alpha\chi_{N.\alpha} +
\lambda_N\chi_{N.N},
\label{hred}\end{equation}
 $\lambda^{...}$ are the Lagrange multipliers, and
\begin{equation}
u_{ij}\equiv{\bf e}_i{\bf e}_j-\delta_{ij},\quad
 \phi_{a.a+1}\equiv {\widetilde\phi}_{a.a+1}-F_{,a},\quad
{\phi}_{0.1} \equiv{\widetilde\phi}_{0.1}+\sum_{i}{\tilde k}_iF_{,i}-F,
\end{equation}
Stabilization of primary constraints (\ref{erunda})
 produces the (secondary) first-stage constraints
\begin{eqnarray}
 & u_{ij}\approx 0;\quad
 s\phi_{0.1}+\sum_ak_a\phi_{a.a+1}\approx 0,  \quad
\Rightarrow {\cal H}\approx 0 ;&\label{u}\\
&s{\phi}_{a.a+1}=-F_{Na}(k_N-2\lambda F_{,N});\quad
 (k_N-2\lambda F_{,N})F_{NN}\approx 0.& \label{primary0}
\end{eqnarray}
$\;$
Now, we can reduce the initial Hamiltonian system by 
the constraints (\ref{erunda}),  and consider the system
with
the symplectic structure
\begin{equation}
\omega_N=d{\bf p}\wedge d{\bf x}+
\sum_{i=1}^N d{\bf p}_i\wedge d{\bf e}_i
\label{ss}\end{equation}
and the Hamiltonian   (\ref{hred}), where
 the expressions (\ref{chi}) and (\ref{u}) define
 the primary constraints. The equations
 (\ref{primary0}) and (\ref{PhiNN0})
either determine variables $ k_a, k_N$ as a functions of
${\widetilde\phi}_{0.1}, {\widetilde\phi}_{a.a+1}$,
 or define a primary constraints, at which the variables  $k_a, k_N$
 represent Lagrange multipliers.
The number of primary constraints, arising in that
way, is equal to  the corank of  $F_{ij}$.  \\

Note that the functions (\ref{sectilde})  form,
with respect to (\ref{ss}), a closed  algebra,
and obey the equations
$$\{{\widetilde\phi}_{...},u_{...}\}
\approx\{{\widetilde\Phi}_{...}, u_{...}\}\approx 0 .$$
The constraints $u_{N.N}, u_{N,\alpha}$
 and $\chi_{N.\alpha}, \chi_{N.N}$ are of the second-class,
$$
\{{\chi}_{N.i},u_{jk}\}\approx\delta_{Nj}\delta_{ik,}
\Rightarrow \lambda_{N\alpha}=\lambda_N=0;
$$
while  the  constraints $u_{N.N-1}$, $u_{a.b}$ are  of the first-class,
 and their stabilization does not generate secondary constraints;
rather, they generate trivial gauge transformations. \\
Consequently, all the secondary constraints are the functions 
of (\ref{sectilde}).      \\

From this follows, that the dimension of 
the phase space of the system, $D_{red}$ satisfy unequality
$$ (2D-3N-2)(N+1)\leq D_{red} \leq  (2D-N)(N+1)-2, $$
where the upper limit corresponds to nondegenerate case, $det F_{ij}\neq0$.

Since  the  gauge transformations of a
system are defined by the primary first-class 
constraints \cite{tyutin}, we  conclude, that 
the number of gauge degrees of freedom of the generalized 
rigid particles does not exceed $corank\;F_{ij}+1$.
 For instance, in a maximally nondegenerate case $det F_{ij}\neq 0$,
 the  Lagrangian possesses only reparametrization invariance.
The system possesses only primary constraints,
and the dimension  of the phase space of that system is 
 $D_{max}=(2D-N)(N+1)-2$. \\

{\bf Example.} The simplest example of nondegenerate  system is 
defined by the Lagrangian
$$F=\frac{1}{2}\sum_{i}^N b_i{\tilde k}^2_i
+\sum_{i=1}c_i{\tilde k}_i +c_0,
\quad b_1\cdot b_2\cdot\ldots\cdot b_N\neq 0.$$

Solving the constraints  (\ref{PhiNN0}) and (\ref{primary0}),  we find
the expressions for curvatures,
\begin{equation}
{\tilde k}_a=({\widetilde\phi}_{a.a+1}-c_a)/b_a,\quad
(b_N{\tilde k}_N+c_N)^2={\widetilde\Phi}_{N.N},
\end{equation}
and the Hamiltonian
 \begin{equation}
   {\cal H}=s{\phi}_{0.1} +d_{ij}u_{ij},\quad
\phi_{0.1}={\widetilde\phi}_{0.1}+\frac 12\sum_{i}b_i{\tilde k}^2_i-c_0.
\end{equation}
The system possesses the following complete set of (primary) constraints
$${\phi}_{0.1}\approx 0,\quad u_{ij}\approx 0,\quad\chi_{N.N}\approx 0,\quad
\chi_{N.\alpha}\approx 0.$$

\setcounter{equation}0
 \section{Lagrangians, Linear on Curvatures}

Let consider the models with maximal set of primary constraints, i.e.
 when $rank F_{ij}=0$.  
In this case the Lagrangians  
are  {\it linear} functions of the  external curvatures,
   \begin{equation}
    F=c_0+ \sum_{i=1}^{N}c_i{\tilde k}_i,
\label{lagrlin}\end{equation}
and can be considered as a potential candidates 
on the role of the systems with maximal gauge degrees of freedom.

Such systems possess  the following set of primary constraints
\begin{eqnarray}
&\phi_{0.1}={\bf p}{\bf e}_1 -c_0\approx 0,&\nonumber\\
& \phi_{a.a+1}={\bf p}_{a}{\bf e}_{a+1}-
{\bf p}_{a+1}{\bf e}_{a} -c_{a}\approx 0,&\nonumber\\
&\Phi_{N.N}={\bf p}_N{\hat L}{\bf p}_N-c^2_N\approx 0,&\nonumber\\
&\chi_{N.N}={\bf p}_N{\bf e}_{N}\approx 0,\quad
\chi_{N.\alpha}={\bf p}_N{\bf e}_{\alpha}\approx 0,&\nonumber\\
&u_{ij}={\bf e}_i{\bf e}_j-\delta_{ij}\approx 0,  & \nonumber\\
\label{primarylinear}\end{eqnarray}
and the Hamiltonian
\begin{equation}
{\cal H}=s\phi_{0.1}+\sum_{a=1}^{N-1}k_a\phi_{a.a+1}+
\lambda\Phi_{N.N}+\sum_{i,j=1}^Nd^{ij}u_{ij}.
\label{slin}\end{equation}
From the equations of motion for ${\bf e}_N$
we can see that $2c_N\lambda=k_N=s{\tilde k}_N$, i.e.
 all the reparametrization invariants 
 play the role  of Lagrange multipliers.

Performing the Legendre transformation we have required the 
condition $k_N\neq 0$. So, stabilizing constraints 
we should suppose $$ k_a\neq 0, \lambda\neq 0. $$

Let impose the gauge conditions ,
 fixing $d_{N.N-1}$ and $d_{a.b}$,
\begin{equation}
\chi_{N.N-1}\equiv{\bf p}_{N}{\bf e}_{N-1}\approx 0,
\quad \chi_{a.a-\kappa}\equiv{\bf p}_{a}{\bf e}_{a-\kappa}\approx 0,
\quad \kappa=0,\ldots,a-1\;,\label{gaugeNN-1}
\end{equation}
which turns all the functions ${\widetilde\Phi}_{i.j}$
  to the quadratic form.
These conditions, 
together with constraints (\ref{chi}), satisfy equations,
$$
 \{\chi_{i.j}, {\widetilde\Phi}_{N.N}\}\approx 2\delta_{N.i}\delta_{ij}
 {\widetilde\Phi}_{N.N};\quad
\{\chi_{i.j}, {\widetilde\phi}_{a.a+1}\}\approx
 \delta_{ij}(\delta_{i.a}-\delta_{i.a+1}){\widetilde\phi}_{a.a+1},$$
$$\;$$
which lead to  the following gauge fixing
\begin{equation}
d_{i.j}= \delta_{ij}(k_ic_i-k_{i-1}c_{i-1}-s\delta_{1.i} c_0).
\label{d}\end{equation}
Stabilization of the remaining primary constraints produce the
following secondary first-stage constraints
\begin{equation}
{\widetilde\phi}_{0.2}\approx 0,
\quad{\widetilde\phi}_{\alpha.\alpha+2}\approx 0,
\quad {\widetilde\Phi}_{N.N-1}\approx 0.
\label{secondary}\end{equation}

 One can  easily see  from the expressions for
        the evolution of functions (\ref{sectilde}),
that the further realization of the Dirac procedure essentially
depends from the values of constants $c_0, c_a$,
\begin{eqnarray}
& {\dot\Phi}_{0.0}=-4\lambda\Phi_{0.N}\phi_{0.N},& \nonumber\\
& {\dot\phi}_{0.i}= -k_{i-1}\phi_{0.i-1} +k_{i}\phi_{0.i+1}
+2\lambda\delta_{iN}\Phi_{0.N},&\nonumber \\
& {\dot\Phi}_{0.i}=-s\delta_{1i}\Phi_{0.0}
 -k_{i-1}\Phi_{0.i-1} +k_i\Phi_{0.i+1}-
2\lambda\Phi_{N.\{i}\phi_{0\}.N},&\nonumber \\
& {\dot\phi}_{i.j}=-s\delta_{1[i}\phi_{j].0}
-k_{i-1}\phi_{i-1.j}+k_{i}\phi_{i+1.j}-k_{j-1}\phi_{i.j-1}
 +k_{j}\phi_{i.j+1} +2\lambda\delta_{N[i}\Phi_{j].N},&\nonumber \\
& {\dot\Phi}_{i.j}=-s\delta_{1\{i}\Phi_{j\}.0}
-k_{i-1}\Phi_{i-1.j}+k_i\Phi_{i+1.j}-k_{j-1}\Phi_{j-1.i}
+k_{j}\Phi_{j+1.i} -2\lambda\Phi_{N.\{i}\phi_{j\}.N}.&\nonumber
\end{eqnarray}

Particularly, if  the Lagrangian (\ref{lagrlin})
is conformal- invariant, i.e.  $c_0=0$, then
 stabilization of ${\phi}_{0.1}\approx 0$
 leads to the following set of the first-class constraints
\begin{equation}
{\widetilde\phi}_{0.i}={\bf p}{\bf e}_i\approx 0,\quad
{\widetilde\Phi}_{0.i}\approx {\bf p}{\bf p}_i\approx 0 ,
\quad {\widetilde\Phi}_{0.0}\approx {\bf p}^2\approx 0,
\label{massless}\end{equation}
which  corresponds, in the pseudo-Euclidean space, to the massless case
\footnote{The total momentum  ${\bf P}$
and the rotation generators ${\bf M}^{(2)}$
 of the system are  defined  by the expressions
$$
{ P}^A ={ p}^A,\quad { M}^{(2)AB}={ p}^{[A}{ x}^{B]}
+\sum_{i=1}^N{ p}^{[A }_i { e}^{B]}_i.
$$
}.

Stabilization of the remaining constraints does not touch 
spatial momentum ${\bf p}$ and coordinates ${\bf x}$ of a system
but only specifies its "intrinsic" space, parametrized by 
${\bf e}_i$, ${\bf p}_i$.

\subsection{${ F}=c{\tilde k}_N$}
Consider the special case, when the Lagrangian  is proportional
to  only one higher curvature, $F=c{\tilde k}_N$, or, equivalently,
$c_0=c_1=....c_{N-1}=0$,   $c_N\equiv c\neq0$ .

For this  model, the Dirac procedure generates  
the maximally possible set 
of constraints, all of which are of the first-class,
\begin{equation}
{\widetilde\phi}_{0.i}\approx 0,\quad
{\widetilde\Phi}_{0.i}\approx 0 ,
\quad {\widetilde\Phi}_{0.0}\approx 0,\quad 
{\widetilde\phi}_{i.j}\approx 0,\quad
{\widetilde\Phi}_{i.j}-c^2\delta_{ij}\approx 0.
\label{isospinmax}\end{equation}

So, this system possesses 
 $(N+1)$  degrees of gauge freedom.
The dimension of its phase space is
\begin{equation}
D_{min}=(2D-3N-2)(N+1).
\label{dimbasic}\end{equation}

Let us impose the gauge conditions (\ref{gaugeNN-1})
and introduce the complex coordinates
\begin{equation}
{\bf z}_i=({\bf p}_i+\imath {c}{\bf e}_i)/{\sqrt{2}},
 \quad\omega_2=d{\bf p}\wedge d{\bf x}+\frac{\imath}{c}
\sum_{i}d{\bf z}_i\wedge d{\bar{\bf z}}_i.
\label{complex}\end{equation}

Now the Hamiltonian of a system reads as 
\begin{equation}
{\cal H}=\frac{s}{2c}\left[
\imath{\sqrt{2}}{\bf p}({\bf{\bar z}}_1-{\bf{z}}_1)
+\imath\sum_{a=1}^{N-1}{\tilde k}_a({\bf z}_{a}{\bf{ \bar z}}_{a +1}
-{\bf z}_{a+1}{\bf{ \bar z}}_{a})+
{\tilde k}_N({\bf z}_{N}{\bf{\bar z}}_{N}-c^2)\right].
\end{equation}
The constraints  (\ref{isospinmax}), (\ref{chi}),
 and gauge conditions (\ref{gaugeNN-1}), read as
\begin{equation}
\Phi^0_{i{\bar j}}\equiv {\bf z}_i{\bf{\bar z}}_j- c^2\delta_{ij}\approx 0,
\quad
\Phi^{+}_{i}\equiv{\bf p}{\bf z}_i\approx 0,
\quad\Phi_0\equiv{\bf p}^2\approx 0,\quad
U^{+}_{ij}\equiv{\bf z}_i{\bf z}_j/2\approx 0,
\label{conscomp} \end{equation}
and form the algebra 
\begin{eqnarray}
 & \{\Phi_{i{\bar j}},\Phi_{k{\bar l}}\}=
\imath c(\delta_{i{\bar l}}\Phi_{k{\bar j}}-
\delta_{k{\bar j}}\Phi_{i{\bar l}}),
 \quad
\{\Phi_{i{\bar j}},\Phi^{+}_{k}\}=-ic\delta_{k{\bar j}}\Phi^{+}_i,&
\nonumber\\
&\{\Phi_{i{\bar j}}, U^{+}_{kl},\}=-ic\left(\delta_{k{\bar j}}U^{+}_{il} +
\delta_{l{\bar j}}U^{+}_{ik}\right),\quad
\{\Phi^+_i, U^{-}_{{\bar j}{\bar k}}\}=
ic\delta_{i\{{\bar j}}\Phi^-_{{\bar k}\}}/2,\quad
\{\Phi^{+}_i,\Phi^{-}_{\bar j}\}=ic\delta_{i{\bar j}}\Phi_0&\nonumber\\
&\{\Phi, U^+_{ij}\}=\{\Phi, \Phi_{i{\bar j}}\}=\{\Phi, \Phi^{+}_{i}\}=0,
 \quad
\{\Phi^+_i, U^+_{jk}\}=\{\Phi^+_i, \Phi^+_{j}\}=
\{U^+_{ij}, U^+_{kl}\}=0, &\nonumber\\
& \{U^{+}_{ij},U^{-}_{{\bar k}{\bar l}}\}=
\imath c\delta_{\{i.\{{\bar k}}\Phi_{j\}{\bar l}\}}/4+
\imath c^3\left(\delta_{i{\bar k}}\delta_{j{\bar l}}
+ \delta_{i{\bar l}}\delta_{j{\bar k}} \right)/2& \nonumber
\end{eqnarray}
where $ \Phi^{-}_{i}={\bar\Phi}^{+}_i$, $U^{-}_{ij}={\bar U}^{+}_{ij}$.\\
So,  $U^{\pm}_{ij}$ are   the second-class costraints,
and the remaining ones are of the first-class.\\

From (\ref{dimbasic})  one can  see that if  $D\leq 4$,
the dynamics is nontrivial only
at $D=4, N=1$, and the dimension of phase space of the system coincides with
that of a (3+1)-dimensional massless particle \cite{misha}.
 In this space, it is possible to "spinorize" the  constraints
(\ref{conscomp}) and to carry out covariant quantization of the system
\cite{rr1}.
As it can be seen from (\ref{conscomp}),  similar trick
 can be performed also
for $N>1$ in $(5+1)$--, $(7+1)$--  and   
 $(9+1)$--dimensional spaces, to resolve the 
 part of second-class constraints.

However, it seems most interesting, that the 
constructed set of constraints coincides with the system of $N+1$-pointing
 discreet string \cite{gp}, \cite{fi}.
 
\subsection{N=1, 2}
We have constructed  above the Hamiltonian systems for generalized rigid
particles, which have  maximal and minimal possible 
(for given $N$ and $D$) dimensions of phase spaces.

We have also mentioned, that, even in the case of Lagrangians, linear
 on curvatures,  the presence of curvatures $k_a$
essentially changes the structure of secondary
constraints.
Consequently , such  systems have 
the  phase spaces  of "intermediate"  dimensions
and less gauge symmetries.

Below we illustrate this phenomena on the examples of Lagrangians,
 linear on  curvatures, in $N=1$ and $N=2$ cases.\\

Let us start from $N=1$, $c_0\neq0$ case.
There is  only one secondary constraint
and the condition on the Lagrange multipliers:
\begin{equation}
{\widetilde\Phi}_{0.1}\approx 0,\quad 
s{\widetilde\Phi}_{00}+k_1c_1c_0\approx 0,
\label{au}\end{equation}

Note, that  $\dot{\widetilde\Phi}_{00}=0$, hence
  $p^2=c^2_0-c_0c_1{\tilde k}_1 =const$,
i.e. the trajectory of the system has constant curvature.\\
In pseudo-Euclidean space the last equations corresponds to the
conservation of mass on the given trajectory.

In  complex coordinates (\ref{complex}), where $c\equiv c_1$,
the complete set of constraints can be represented
by one real and two holomorphic constraints
 \begin{equation}
\Phi\equiv{\bf z}{\bar{\bf z}}-c^2_1\approx 0 ,
\quad  U^{+}\equiv {\bf z}^2/{2}\approx 0,
\quad \Phi^+\equiv {\bf p}{\bf z}-ic_0c_1/{\sqrt{2}}\approx 0,
\label{constZ}\end{equation}
forming the algebra
\begin{eqnarray}
 &\{\Phi,\Phi_{+}\}=-\imath c_1\Phi_{+} +c^2_1c_0/{\sqrt 2},
 \quad \{\Phi,U^{+}\}=-2\imath c_1 U^{+},\quad
 \{U^{+},U^{-}\}=\imath c_1\Phi +\imath c_1^3,&\nonumber\\
& \{\Phi^{+},\Phi^{-}\}=\imath c_1{\bf p}^2 ,\quad
 \{\Phi^{+},U^{-}\}=\imath c_1\Phi^- +c_0c^2_1/{\sqrt 2},\quad
 \{\Phi^{+},U^{+}\}=0&\nonumber
\end{eqnarray}
where  $\Phi^{-}\equiv {\bar\Phi^{+}}$, $U^{-}\equiv {\bar U^{+}}$.\\
 So, for $c_0\neq0, N=1$ the dimension of phase space is equal to
$D_{red}=2(2D-3)$.

The system possesses one gauge degree of freedom given by the
Hamiltonian
\begin{equation}
{\cal H}_{1.c_0\neq0}=\frac{s}{2c^2_1c_0}
\left[ 2\imath c^2_1c_0 {\bf p}({\bf z}-{\bf{\bar z}}) +
(c^2_0 -{\bf p}^2)({\bf z}{\bar{\bf z}}-c^2_1)\right].
\end{equation}

At $c_0=0$ (see Subsection 3.1)
the dimension of system phase space equals to $D_{red}=2(2D-5)$
and it has two gauge degrees of freedom.\\

%

Now let us consider the case $N = 2$ with arbitrary constants
$c_0, c_1, c_2$.

When $c_0\neq0$, we have two secondary
second-class constraints  defined
by expressions (\ref{secondary}).
 Their stabilization results in the conditions
\begin{equation}
c_2c_0k_1=k_2{\widetilde\Phi}_{0.2},\quad 
sc_0c_2{\widetilde\Phi}_{0.2}=k_2\left({\Phi}_{1.1}
{\widetilde\Phi}_{0.2}-c^2_2c_1c_0\right),
\end{equation}
  where $\Phi_{1.1}\equiv{\widetilde\Phi}_{1.1}-c^2_2$,
and ${\widetilde\Phi}_{0.2}\neq 0 $.\\
At $c_1=0$ the second condition takes the form
$sc_0c_1+ k_2\Phi_{1.1}=0$.\\
The system possesses one gauge degree of freedom. The dimension of its
phase space equals  to $D_{red}=6(D-2)$.
Like to the case  of $N=1$, there is the motion constant, 
${\widetilde\Phi}_{0.0}$,
 which corresponds in the pseudo-Euclidean space,
to the   conservation of the mass on a given trajectory.   \\

When  $c_0=0, c_1\neq 0$,
the secondary constraints are defined by expressions
(\ref{massless}) and ${\widetilde\Phi}_{1.2}\approx 0$.
There is the condition
\begin{equation}
k_2c_2c_1+k_1\Phi_{1.1}=0.
\end{equation}      
Notice, that  ${\Phi}_{1.1}$ is  the motion constant 
${\dot\Phi}_{1.1}=0$, so ${\tilde k}_2/{\tilde k}_1=const$.\\
This system possesses two gauge degrees of
 freedom. The dimension of its phase space is 
$D_{\rm red}=2(3D-10)$.     \\

At $c_0=c_1=0$ (see Subsection 3.1),
 the dimension of phase space of the system
equals $D_{red}=6(D-4)$, and there are three gauge degrees of freedom.

\section{Conclusion.}
We presented
 the Hamiltonian formulation for the models of generalized 
rigid particles, based on the 
use of the  moving frame.
This strongly simplify  formulation of  the  system  and its subsequent 
 analyses.
In particular, we found that the dimension of phase space 
of the system
with the Lagrangian depending on the first $N$ external curvatures satisfy
 the inequality
$$ (2D-3N-2)(N+1)\leq D_{red}
\leq  (2D-N)(N+1)-2, $$
where the upper limit is corresponds to the Lagrangian, quadratic
 on first $N$ curvatures, while the lower
limit  corresponds  to the  Lagrangian proportional to 
 $N$-th curvature. 

In the first case, the Lagrangian possess 
only reparametrization 
degree of freedom, while in the last case, it has
 $(N + 1)$ gauge degrees of freedom.
Moreover, in the last case, in appropriate gauge  fixing, the
complete set of constraints and gauge-fixing conditions
become quadratic, and coincides with   the  $N+1$-particle discreet
 string \cite{gp}, \cite{fi}, which was quantized recently
 in BRST approach  both  for  $N=1$ \cite{pt1}
  as well as for arbitrary $N$ \cite{ptN}.
We think, that this surprising parallel deserve to be studied separately.

In the case of Lagrangian with arbitrary linear dependence
from curvatures, the set of primary 
constraints turns out to be quadratic too.
However, the full set of secondary constraints  is essentially depending 
by the constants $c_i$, although 
the algorithm of  constructing the secondary
constraints, and the generators of gauge symmetries, is the sequence
 of algebraic operations.

\section{Acknowledgments.}

The author is  grateful to  E. Ivanov for suggestion to study
the Lagrangians, depending on curvatures, 
for the explanation of the importance of this problem
in the context of $W-$ algebras,   useful discussions and criticism.

Several useful discussions, concerning Dirac's procedure,
 took place with P. Pyatov and S. Lyakhovich,
and on spinning particles with S. Lyakhovich, 
A. Sharapov, and, especially, K. Shekhter.
A. Pashnev turn my attention to the intriguing 
correspondence between rigid particles  and discrete strings, 
and gives the references of his works on the BRST quantization.
  This work was finished also due to permanent 
interest and useful remarks of O. Khudaverdian, S. Krivonos and 
V. Nesterenko.
All of them I  express my deep gratitude .

I would like to acknowledge  I. V. Tyutin,
who kindly explain me, in grate details, the structure of
 gauge transformations in constraint Hamiltonian 
systems. 

I am thankful  to  S. Randjbar- Daemi for kind hospitality
at the International Center of Theoretical Physics, Trieste, Italy,
where the work has been completed.\\

The work has been partially supported by grants INTAS-RFBR No.95-0829,
INTAS-96-538 and INTAS-93-127-ext.

\end{document}